\documentclass[%
 amsmath,amssymb,
 aps,
 prd,
twocolumn, superscriptaddress
]{revtex4-1}

\usepackage[normalem]{ulem}
\usepackage{cancel}

\usepackage[dvipsnames]{xcolor}
\usepackage{graphicx}
\usepackage{dcolumn}
\usepackage{bm}
\usepackage{xcolor}
\usepackage{orcidlink}
\hypersetup{colorlinks, citecolor=green}
\hypersetup{colorlinks=true, linkcolor=OrangeRed, urlcolor=MidnightBlue}
\usepackage{hyperref}
\usepackage{footnote}

\usepackage{empheq} 
\usepackage{color} 
\definecolor{lightgray}{gray}{0.91}

\begin{document}


\title{Normal oscillation modes and radial stability of neutron stars with a dark-energy core from the Chaplygin gas}

\author{Juan M. Z. Pretel \orcidlink{0000-0003-0883-3851}}
 \email{juanzarate@cbpf.br}
 \affiliation{Centro Brasileiro de Pesquisas F{\'i}sicas, Rua Dr.~Xavier Sigaud, 150 URCA, Rio de Janeiro CEP 22290-180, RJ, Brazil
}

\author{Mariana Dutra \orcidlink{0000-0001-7501-0404}}
 \email{marianad@ita.br}
 \affiliation{Departamento de F{\'i}sica, Instituto Tecnol{\'o}gico de Aeron{\'a}utica, DCTA, 12228-900, S{\'a}o Jos{\'e} dos Campos, SP, Brazil
}
 \affiliation{Université de Lyon, Université Claude Bernard Lyon 1, CNRS/IN2P3, IP2I Lyon, UMR 5822, F-69622, Villeurbanne, France
}

\author{Sergio B. Duarte \orcidlink{0000-0003-0027-0230}}
 \email{sbd@cbpf.br}
 \affiliation{Centro Brasileiro de Pesquisas F{\'i}sicas, Rua Dr.~Xavier Sigaud, 150 URCA, Rio de Janeiro CEP 22290-180, RJ, Brazil
}

\date{\today}

\begin{abstract}
As a potential candidate for the late-time accelerating expansion of the Universe, the Chaplygin gas and its generalized models have significant implications to modern cosmology. In this work we investigate the effects of dark energy on the internal structure of a neutron star composed of two phases, which leads us to wonder: Do stable neutron stars have a dark-energy core? To address this question, we focus on the radial stability of stellar configurations composed by a dark-energy core --- described by a Chaplygin-type equation of state (EoS) --- and an ordinary-matter external layer which is described by a polytropic EoS. We examine the impact of the rate of energy densities at the phase-splitting surface, defined as $\alpha= \rho_{\rm dis}^-/\rho_{\rm dis}^+$, on the radius, total gravitational mass and oscillation spectrum. The resulting mass-radius diagrams are notably different from dark energy stars without a common-matter crust. Specifically, it is found that both the mass and the radius of the maximum-mass configuration decrease as $\alpha$ becomes smaller. Furthermore, our theoretical predictions for mass-radius relations consistently describe the observational measurements of different massive millisecond pulsars as well as the central compact object within the supernova remnant HESS J1731-347. The analysis of the normal oscillation modes reveals that there are two regions of instability on the $M(\rho_c)$ curve when $\alpha$ is small enough indicating that the usual stability criterion $dM/d\rho_c>0$ still holds for rapid phase transitions. However, this is no longer true for the case of slow transitions.
\end{abstract}

\maketitle


\section{Introduction}

The mysterious component that leads to an accelerated expansion phase of the Universe has been attributed to several candidates, such as: A dynamical scalar field in quintessence models \cite{Ratra1988, Joyce2016}, a single scalar field interacting with gravity by means of non-canonical kinetic terms in the so-called $k$-essence models \cite{Armendariz1999, Chiba2000}, modified gravity theories \cite{Joyce2016, Shankaranarayanan2022}, vacuum energy predicted by the quantum field theory \cite{Weinberg1989}, extra dimensions \cite{Montefalcone2020, Jalalzadeh2023}, an exotic form of fluid with large negative pressure that describes the transition from a Universe filled with dust-like matter to an exponentially expanding Universe in the late times \cite{Kamenshchik2001}, among other models proposed in the literature. The well-known $\Lambda$CDM model, where $\Lambda$ is the cosmological constant, is the standard model of contemporary cosmology and the simplest model to describe dark energy. Such a model is in excellent agreement with recent observational data \cite{Aghanim2020}, however, it suffers from two problems \cite{Weinberg1989, Padmanabhan2003}: (i) Fine-tuning problem, where the current energy density of the cosmological constant $\rho_{\Lambda} \sim 10^{-47}\, \rm GeV$ \cite{Padmanabhan2003, Copeland2006} is in conflict with the value of vacuum energy density $\rho_{vac} \sim 10^{74}\, \rm GeV$ \cite{Weinberg1989}, and hence $\rho_\Lambda$ requires fine-tuning. (ii) The matter-energy density $\rho_m$, which changes with time, and the dark energy density (staying constant) are of the same order today. This is known as the cosmic coincidence problem, and would require a correction of the parameters in the early epoch of the Universe.

One of the candidates for dark energy aforementioned in the first paragraph is the Chaplygin gas \cite{Kamenshchik2001}, which offers a simple phenomenological description and provides a plausible solution for the unification of two uncharted components; dark matter and dark energy \cite{BILIC2002, Gorini2003, Xu2012}. An interesting feature of this model is that it behaves like a cosmological constant at late stage (or, lower redshifts) and as dust-like matter (pressureless fluid) at early stage. Furthermore, in the light-cone parameterization, the original Chaplygin gas model can be obtained from the string Nambu–Goto action for $d$-branes moving in a $(d+2)$-dimensional spacetime \cite{Bordemann1993, Ogawa2000, Jackiw2000}. Kamenshchik \textit{et al.}~\cite{Kamenshchik2001} also introduced a generalized Chaplygin gas, given by $p= -B/\rho^\alpha$, where $B$ and $\alpha$ are real constant parameters and $\rho$ is the energy density. The cosmological aspects of this model for the Universe evolving from a phase dominated by nonrelativistic matter to a phase dominated by a cosmological constant were examined in Ref.~\cite{Bento2002}. In fact, such a model has been confronted with different sets of observational data \cite{Xu2012, Bento2003, Bertolami2004, Barreiro2008}. An even more generalized version was proposed by Zhang and colleagues \cite{Zhang2006}, where the constant $B$ becomes a function depending on the scale factor. See Ref.~\cite{Mamon2022} for a more recent study by assuming this type of generalization.

Over the years, the original Chaplygin gas model has gone through some modifications and in its most generalized form it is characterized by an EoS of the form $p= A\rho - B/\rho^\alpha$, where the three degrees of freedom $\{A,B,\alpha\}$ have been measured by means of the \textit{Planck} 2015 CMB anisotropy, type-Ia supernovae and observed Hubble parameter data sets \cite{Li2019}. Yang and collaborators \cite{Yang2019} have explored whether the global $21$-cm absorption signal detected by EDGES can improve our understanding of the Chaplygin gas models. Indeed, it was found that the uncertainties on the parameters of the Chaplygin gas models can be reduced by a factor between $1.5$ and $10$. In addition, based on latest observations of high-redshift quasars, a series of Chaplygin gas models as candidates for dark matter-energy unification were recently investigated by Zheng \textit{et al.}~\cite{Zheng2022}.

If we assume that dark energy is an exotic fluid responsible for the accelerated expansion of the Universe, then it must be present in any region of spacetime, including spherically symmetric objects such as compact stars \cite{Lobo2006, Gorini2008PRD}. In that regard, the effects of dark energy (described by a Chaplygin-type EoS) on the relativistic structure of single-phase compact stars have been investigated over the last few years \cite{Rahaman2010, Bhar2018, Panotopoulos2020EPJP, Tello2020, Panotopoulos2021, Estevez2021, Prasad2021, Pretel2023EPJC, Sunzu2023, Kumar2023}. In fact, for such an EoS, it has been shown that the Tolman-Oppenheimer-Volkoff (TOV) equations provide maximum masses above $2\, M_\odot$ \cite{Panotopoulos2020EPJP, Panotopoulos2021, Pretel2023EPJC}, which favors the observational measurements. These dark-energy stellar configurations form a mass-radius diagram similar to that of quark matter, obey the causality condition, and are dynamically stable under radial perturbations when $dM/d\rho_c>0$ on the $M(\rho_c)$ curve \cite{Panotopoulos2020EPJP, Pretel2023EPJC}.

The findings reported in Ref.~\cite{Gorini2008PRD} were extended a year later assuming a generalized Chaplygin gas \cite{Gorini2009PRD}, where the authors addressed the superluminality issue. Specifically, they investigated how a modification of the EoS $p= -\Lambda^{\alpha+1}/\rho^\alpha$, required by causality arguments at densities very close to $\Lambda$, affects the previous results obtained in \cite{Gorini2008PRD}. We should point out that the literature also offers the study of wormholes constructed by means of the Chaplygin gas. As a matter of fact, Eiroa and Aguirre \cite{Eiroa2012} constructed spherically symmetric thin-shell wormholes supported by a generalized Chaplygin gas in Born-Infeld electrodynamics coupled to Einstein gravity. Furthermore, Kuhfittig \cite{Kuhfittig2015} discussed a natural way to obtain a complete wormhole solution by considering that the wormhole is supported by generalized Chaplygin gas and admits conformal Killing vectors. All these investigations have been carried out under the assumption that the star-like objects are described only by an EoS corresponding to the Chaplygin gas. In the present work we are interested in a hybrid context, where the dark energy is confined to the core of the compact star but there is also a crust described by another EoS.

Under a Newtonian formalism, polytropic models have been recently investigated by considering a generalized Chaplygin EoS \cite{Abellan2023}, where the authors provide the Lane-Emden equation for particular cases of the anisotropy factor. It is also worth commenting that, within the framework of massive gravity, the effect of the massive graviton on dark energy star structure was analyzed by Tudeshki \textit{et al.}~\cite{Tudeshki2023}. For further compact-star models with dark energy in modified gravity we also refer the reader to Refs.~\cite{Tudeshki2023PLB, Bhar2023PDU, Bhar2023FP, DAS2024}. Motivated by these studies, we will extend the investigation of single-phase compact stars to a hybrid scenario, but considering dark energy as described by the modified Chaplygin gas. Therefore, a crucial open question, to which this work aims to contribute, concerns the confinement of dark energy in the core of a stable compact star and adding an outer layer (or crust) on the core. Of course, a rigorous analysis of the stability of these stars involves examining the normal oscillation modes when they are adiabatically perturbed. In addition, we will investigate whether the observational mass-radius measurements can be explained with an ordinary matter EoS for the crust and a Chaplygin-like EoS for the core of the hybrid star.

The organization of the present work is as follows: In Sec.~\ref{Sec2} we briefly summarize the hydrostatic equilibrium of compact stars in Einstein gravity. In the same section we also present the EoSs describing both phases and discuss the parameter space for polytropic stars with a dark-energy core from the Chaplygin gas. Section \ref{Sec3} deals with radial perturbations and the stability of two-phase relativistic compact stars, as well as presents the junction conditions at the phase-splitting interface for slow and rapid transitions. In Sec.~\ref{results} we discuss our numerical results based on the mass-radius diagrams and oscillation spectrum, and finally, our conclusions are presented in Sec.~\ref{conclusion}.


\section{Static equilibrium configurations}\label{Sec2}

Since our compact-star model is a hybrid star, here we present the basic stellar structure equations and briefly describe the two-phase fluid by specifying the EoS for both the dark-energy core and the ordinary-matter crust.

\subsection{TOV equations}

To investigate the normal vibration modes of compact stars in the presence of dark energy, we first need to obtain the background solutions given a spacetime metric and a matter-energy distribution. In general relativity, the spacetime curvature and the energy-momentum content are related by the Einstein field equations, namely
\begin{equation}\label{FieldEq}
    R_{\mu\nu} - \frac{1}{2}Rg_{\mu\nu} = 8\pi T_{\mu\nu} ,
\end{equation}
where $R_{\mu\nu}$ is the Ricci curvature tensor, $R$ is the Ricci scalar, $T_{\mu\nu}$ the energy-momentum tensor, and $g_{\mu\nu}$ is the metric tensor which determines the invariant square of an infinitesimal line element. The equilibrium star is assumed to be composed of two different layers of isotropic perfect fluids. We now proceed to consider spherically symmetric stellar configurations so that the line element takes the usual form in spherical coordinates
\begin{equation}\label{metricEq}
    ds^2= -e^{2\psi}dt^2 + e^{2\lambda}dr^2 + r^2(d\theta^2+ \sin^2\theta d\phi^2) .
\end{equation}

Besides, the matter-energy distribution of this (hybrid) stellar system is characterized by the energy-momentum tensor $T_{\mu\nu}= (\rho+ p)u_\mu u_\nu+ pg_{\mu\nu}$, where $\rho$ denotes the energy density, $p$ is the pressure and $u^\mu$ is the four-velocity of the perfect fluid. In the hydrostatic equilibrium state we can write $u^\mu= e^{-\psi}\delta_0^\mu$ and $T_\mu^\nu = {\rm diag}(-\rho, p, p, p)$ since $u^{\mu}$ satisfies the normalisation condition $u_\mu u^\mu =-1$. Consequently, the relativistic structure of a unperturbed compact star is described by the TOV equations:
\begin{align}
    \frac{dm}{dr} &= 4\pi r^2\rho ,  \label{TOV1}  \\
    \frac{dp}{dr} &= -(\rho+ p)\left[ \frac{m}{r^2}+ 4\pi rp \right]\left[ 1- \frac{2m}{r} \right]^{-1} ,  \label{TOV2}  \\
    \frac{d\psi}{dr} &= -\frac{1}{\rho+ p}\frac{dp}{dr} ,  \label{TOV3}
\end{align}
that is, a set of three first-order differential equations for the four variables $m$, $\rho$, $p$ and $\psi$. It is important to remark that the metric function $\lambda(r)$ is determined from the relation $e^{-2\lambda}= 1- 2m/r$, where $m(r)$ is a mass function along the radial coordinate.

However, given an EoS of the form $p= p(\rho)$, the number of variables is reduced to three and therefore three boundary conditions are required:
\begin{align}\label{BCforTOV}
    \rho(0) &= \rho_c,  &  m(0) &= 0,  &  \psi(R) &= \frac{1}{2}\ln\left[ 1- \frac{2M}{R} \right] ,
\end{align}
where $\rho_c$ is the central energy density, and $M= m(R)$ is the total gravitational mass of the star calculated at its surface where the pressure vanishes, i.e.~when $p(r=R) =0$ with $R$ being the surface radius. Note that the second condition in Eq.~(\ref{BCforTOV}) demands regularity at the center of the star, while the third condition comes from the continuity of the metric at the surface since the exterior spacetime is described by the Schwarzschild vacuum solution. Because we are dealing with compact stars composed of two phases, separated by a discontinuous surface, we need different EoSs to describe the hybrid system. In the next subsection we will define these EoSs in detail.

\begin{figure}
 \includegraphics[width=8.0cm]{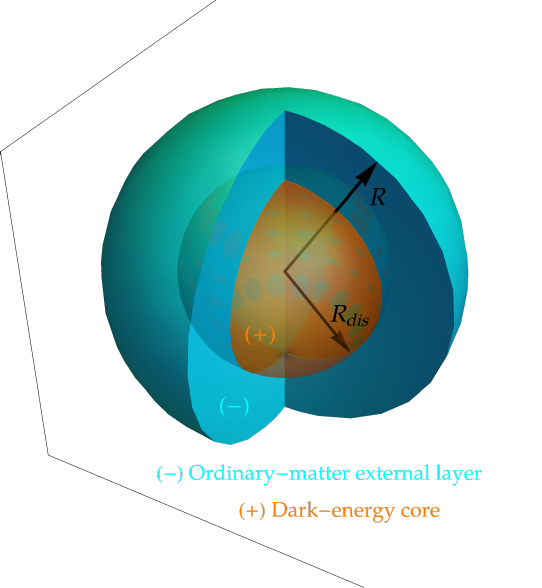}
 \caption{\label{FigHybridStar} Illustrative representation of a hybrid star composed by an ordinary-matter crust and a dark-energy core, where $R_{\rm dis}$ indicates the radius of the discontinuous surface and $R$ is the surface radius where the pressure is zero. Actually, this can be interpreted as confining the dark energy to the core of the compact star, with the crust surrounding it. }  
\end{figure}

\begin{figure}
 \includegraphics[width=8.5cm]{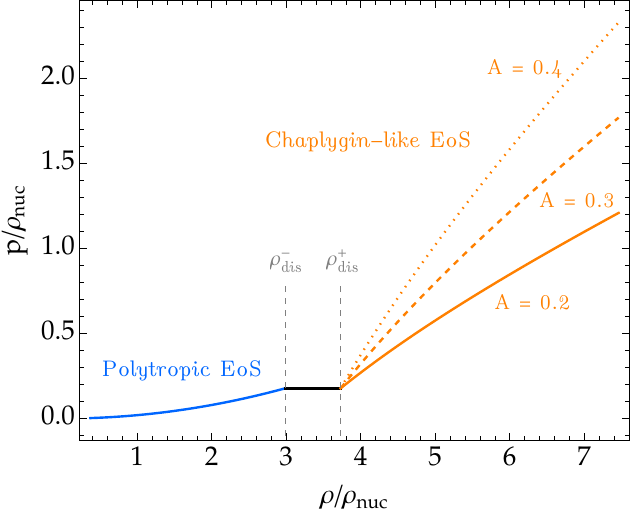}
 \caption{\label{FigEoSs} Some equations of state with energy-density discontinuity for hybrid stars with dark energy and ordinary matter phases. Both energy density and pressure are normalized by the standard nuclear density, i.e.~$\rho_{\rm nuc}= 2.68\times 10^{14}\, \rm g/cm^3$. Three different values of $A$ were considered and we have also used $\rho_{\rm dis}^+= 1.0 \times 10^{15}\, \rm g/cm^3$, $\alpha= 0.8$, $\eta= 1.0$, $\kappa= 100\, \rm km^2$ and central density $\rho_c= 2.0 \times 10^{15}\rm g/cm^3$. The values adopted for the different parameters (\ref{Parameters}) in this plot have been considered for illustration purposes only. }  
\end{figure}

\subsection{Equations of state}

The stellar structure equations have to be supplemented with an EoS of the form $p = p(\rho)$, namely, a functional relation between the energy density $\rho$ and pressure $p$ inside the stellar fluid. Here, we consider a compact star composed of two phases: a dark-energy core described by a Chaplygin-type EoS, and an outer layer of ordinary matter described by a polytropic EoS. Therefore, the EoS is given by
\begin{equation}\label{EoSEq}
    p(\rho) = 
  \begin{cases}
    A\rho- \frac{B}{\rho},  & \quad  0 \leq r \leq R_{\rm dis} ,  \\ 
    \kappa\rho^{1+ 1/\eta},  & \quad  R_{\rm dis} \leq r \leq R ,
  \end{cases}
\end{equation}
with $R_{\rm dis}$ being the radius of the discontinuous surface (where the pressure corresponds to the energy-density discontinuity) and $R$ denotes the stellar radius (where the pressure vanishes). The extra term ``$-B/\rho$'', with $B$ being a positive constant (given in $\rm m^{-4}$ units), represents a negative pressure that leads to the accelerated expansion of the current Universe \cite{Kamenshchik2001}. The quintessence model fails to avoid fine-tuning in explaining the cosmic coincidence problem, however, it was shown that the Chaplygin gas provides a good alternative to explain the transition from a Universe filled with dust-like matter to an accelerated expansion phase \cite{Zheng2022}, although today there are more generalized versions as we discussed in the introduction. Moreover, this negative pressure term has a well-defined connection with string and brane theories \cite{Kamenshchik2001, Ogawa2000}. Meanwhile, the contribution ``$A\rho$'' describes a barotropic fluid, where $A$ is a positive dimensionless constant. We are taking this linear term into consideration in order to maintain an EoS that leads to compact stars with dark energy, as adopted by other researchers in the case of single-phase compact stars \cite{Rahaman2010, Bhar2018, Panotopoulos2020EPJP, Tello2020, Panotopoulos2021, Estevez2021, Pretel2023EPJC, Sunzu2023}. It has further been argued that small values of $A$ lead to small maximum masses which are not consistent with the current observational measurements \cite{Pretel2023EPJC}. It is important to emphasize that $A$ has to assume values such that they satisfy the causality condition along the entire radial coordinate inside the star.

Following the notation convention adopted in Ref.~\cite{Sotani2001}, hereafter, we use ``$+$'' and ``$-$'' to label the energy densities in the regions $r< R_{\rm dis}$ and $r> R_{\rm dis}$, respectively. See Fig.~\ref{FigHybridStar} for a graphical representation of the compact star with a dark-energy core and the crust surrounding it. The transition pressure occurs at the discontinuity surface, which implies that we have an inner energy density $\rho_{\rm dis}^+$ and an outer energy density $\rho_{\rm dis}^-$. Since the pressure must be continuous at $r= R_{\rm dis}$, one can obtain an explicit expression for $B$ in terms of the other parameters, that is, 
\begin{equation}\label{EqB}
    B = A(\rho_{\rm dis}^+)^2 - \kappa (\rho_{\rm dis}^+)(\rho_{\rm dis}^-)^{1+ 1/\eta} ,
\end{equation}
where, of course, $\rho_{\rm dis}^- \leq \rho_{\rm dis}^+$. It is pertinent to introduce a new parameter defined as the ratio of the outer density to the inner density at the discontinuous surface, i.e., $\alpha = \rho_{\rm dis}^-/\rho_{\rm dis}^+ \leq 1$. Of course, the particular case $\alpha= 1$ means that the density is continuous. As a consequence, our stellar model is characterized by a set of six free parameters:
\begin{empheq}[box=\fcolorbox{cyan}{lightgray}]{equation}\label{Parameters}
      \left\lbrace \rho_c,\ \rho_{\rm dis}^+,\ \alpha,\ \eta,\ \kappa,\ A \right\rbrace
\end{empheq}
In Fig.~\ref{FigEoSs} we illustrate the EoS (\ref{EoSEq}) with discontinuous density for some specific values of the above parameters. In the remaining part of this work, for the polytropic EoS, we will establish $\eta= 1.0$ and $\kappa= 100\, \rm km^2$, which are typical values to describe neutron stars \cite{Allen1998, KokkotasRuoff2001, PretelJCAP2021}.


\section{Radial pulsation equations}\label{Sec3}

Since we are dealing with extremely high energy densities, the general relativistic effects become important even when a compact star is subjected to radial perturbations. Chandrasekhar pioneered the radial stability of single-phase relativistic compact stars (either quark or hadronic matter) \cite{ChandrasekharApJ, ChandrasekharPRL}, and since then the radial oscillation equations have been written in different forms for numerical convenience, see for example Refs.~\cite{Chanmugam1977, Benvenuto1991, VathChanmugam, Gondek1997, KokkotasRuoff2001, Pretel2020MNRAS, Bora2021, Hong2022, Hong2023}. In this study we will deal with the dynamical stability of compact stars containing two phases, however, each phase is described by the equations already known for the homogeneous case. The equilibrium solutions provided by the TOV equations (\ref{TOV1})-(\ref{TOV3}) describe a family of stellar configurations, and to test their stability towards gravitational collapse or explosion we must calculate the frequencies of normal vibration modes. These frequencies can be found by considering small deviations from the hydrostatic equilibrium state, so that a linear theory can be applied to the Einstein field equations. Furthermore, the radial pulsations are assumed to be harmonic and adiabatic, so that the fluid elements of the star neither gain nor lose heat during the vibration. 

The linearized perturbation equations can be obtained by introducing the Lagrangian displacement $\xi$ around the equilibrium position, namely, the fluid element located at $r$ in the unperturbed system is displaced to the position $r+ \xi(t,r)$ in the perturbed system. Thus, we can write $\xi(t,r)= \chi(r)e^{i\omega t}$, where $\chi(r)$ and $\omega$ are the amplitude and the vibration frequency of the standing wave, respectively. Defining $\zeta= \chi/r$, the adiabatic radial oscillations of relativistic stars are governed by the following first-order time-independent equations
\begin{align}
    \dfrac{d\zeta}{dr} =& -\dfrac{1}{r}\left( 3\zeta + \dfrac{\Delta p}{\gamma p} \right) + \dfrac{d\psi}{dr}\zeta ,  \label{ROEq1} \\
    \dfrac{d(\Delta p)}{dr} = &\ \zeta\left[ \omega^2e^{2(\lambda - \psi)}(\rho + p)r - 4\dfrac{dp}{dr} \right.  \nonumber   \\
    & \left.- 8\pi e^{2\lambda}(\rho+ p)rp + r(\rho+ p)\left( \dfrac{d\psi}{dr} \right)^2  \right]   \nonumber  \\
    & - \Delta p\left[ \dfrac{d\psi}{dr} + 4\pi (\rho +p)re^{2\lambda} \right] ,  \label{ROEq2}
\end{align}
where $\gamma = (1 + \rho/p)dp/d\rho$ denotes the adiabatic index at constant specific entropy, and $\Delta p = \delta p + \chi dp/dr$ is the Lagrangian perturbation of the pressure, with $\delta p$ being the Eulerian perturbation. Note that all metric and fluid quantities in the radial oscillation equations are determined from the static background. 

It is evident that Eq.~(\ref{ROEq1}) has a singularity at the stellar center ($r=0$). Thus, to guarantee regularity we must demand that
\begin{equation}\label{BCforRO1}
    \Delta p = -3\zeta\gamma p  \qquad  \text{as}  \qquad  r\rightarrow 0 ,
\end{equation}
and since the surface of the hybrid star is determined by the condition $p(r=R) =0$, we should also impose the following boundary condition for the Lagrangian perturbation of the pressure
\begin{equation}\label{BCforRO2}
    \Delta p = 0  \qquad  \text{as}  \qquad  r\rightarrow R .
\end{equation}

It is important to note that the boundary conditions (\ref{BCforRO1}) and (\ref{BCforRO2}) established at the center and surface of the star, respectively, remain the same as for single-phase stars. Nevertheless, since in this work we are dealing with a two-phase compact star, it becomes necessary to establish junction conditions at the phase-splitting interface for the numerical integration of the oscillation equations (\ref{ROEq1}) and (\ref{ROEq2}). These conditions depend on the velocity of the phase transition near the discontinuous surface. Here we will use the boundary conditions deduced by Pereira \textit{et al.}~\cite{Pereira2018}:

\begin{itemize}
    \item[a)] For slow phase transitions, the volume elements near the discontinuous surface do not change their nature due to the radial perturbations but they co-move with the interface. In such scenario, there is no mass transfer from one phase to another, and the junction conditions at the splitting surface are given by
    \begin{align}\label{JuncCond1}
        \left[ \zeta \right]_-^+ &= 0,  &  \left[ \Delta p \right]_-^+ &= 0,
    \end{align}
    where $[z]_-^+ = z^+- z^-$, with $z$ representing any variable across the interface.

    \item[b)] For rapid phase transitions, there is an instantaneous change in the nature of the volume elements near the interface due to pulsations, and this implies a mass transfer between the two phases. The matching conditions at the interface are as follows
    \begin{align}\label{JuncCond2}
        \left[ \zeta- \frac{\Delta p}{rp'} \right]_-^+ &= 0,  &  \left[ \Delta p \right]_-^+ &= 0,
    \end{align}
    where $p' = dp/dr$ is defined in the hydrostatic equilibrium state.
    
\end{itemize}

\begin{figure*}[htp!]
 \includegraphics[width=5.96cm]{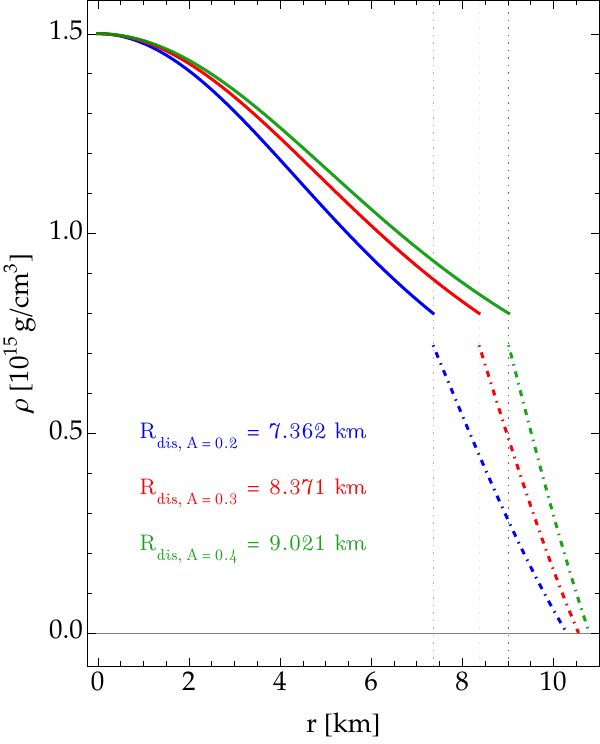}
 \includegraphics[width=5.75cm]{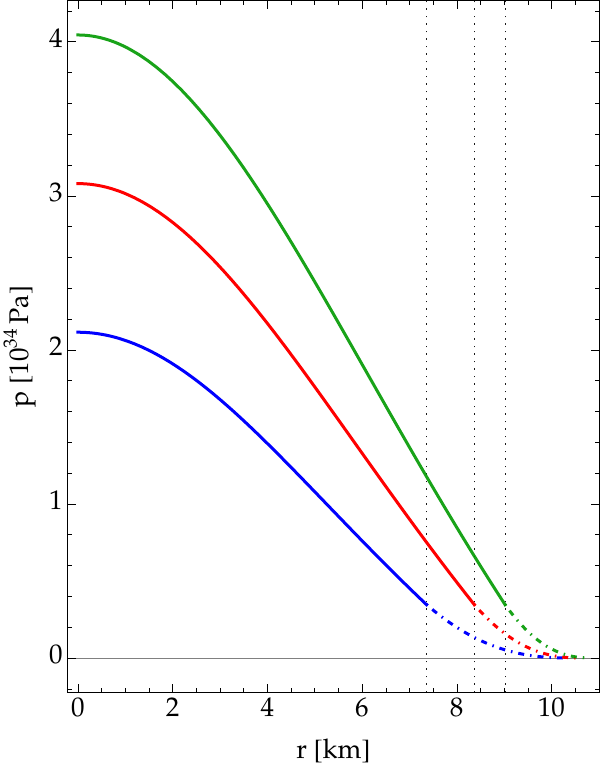}
 \includegraphics[width=5.91cm]{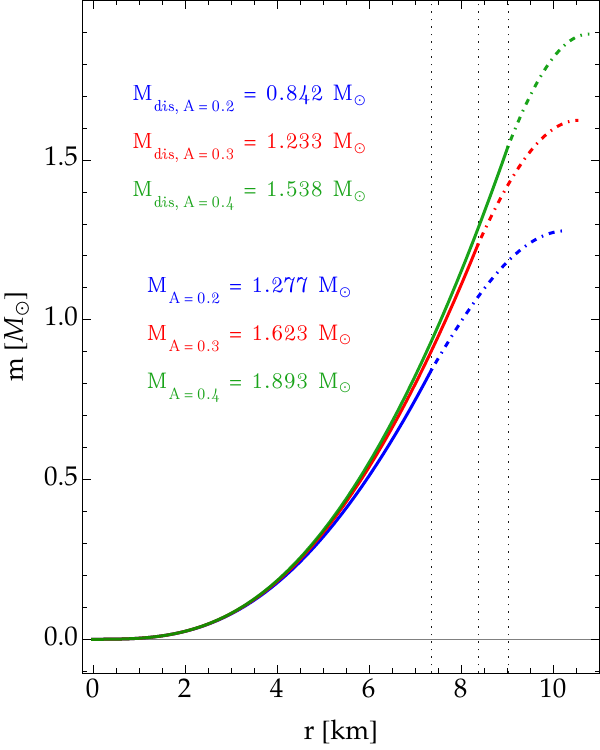}
 \caption{\label{FigOneStar} Radial profile of energy density (left panel), pressure (middle panel), and mass function (right panel) within a stellar configuration with central density $\rho_c= 1.5 \times 10^{15}\rm g/cm^3$, ratio of densities at the discontinuous surface $\alpha= 0.9$, inner energy density $\rho_{\rm dis}^+= 0.8 \times 10^{15}\, \rm g/cm^3$, and three values of the free parameter $A$. In all plots, the solid and dot-dashed curves correspond to the dark energy core and ordinary matter outer layer, respectively. The gray vertical lines indicate the discontinuous surface for each value of $A$, see left plot for the specific values of $R_{\rm dis}$. Remarkably, both the mass of the core and the total mass of the star increase significantly due to the increase in $A$. For this construction and in the next figures, we have adopted $\eta= 1.0$ and $\kappa= 100\, \rm km^2$. }  
\end{figure*}


\section{Numerical results}\label{results}

\subsection{Equilibrium configurations}

Through the first two boundary conditions given in Eq.~(\ref{BCforTOV}), we solve the stellar structure equations (\ref{TOV1}) and (\ref{TOV2}) from the center to the surface of the star. However, since the configuration contains two phases, we follow the steps below:
\begin{itemize}
    \item[$\star$] We integrate from the stellar origin at $r=0$ up to the discontinuity radius at $r= R_{\rm dis}$ (i.e., the radial coordinate corresponding to $\rho= \rho_{\rm dis}^+$) with Chaplygin-like EoS.
    \item[$\star$] We then integrate from the interface at $r= R_{\rm dis}$ up to the surface of the star at $r= R$ (this is, the radial coordinate where the pressure vanishes) with polytropic EoS.
\end{itemize}
In particular, given the EoS (\ref{EoSEq}), in Fig.~\ref{FigOneStar} we display the numerical solution for an equilibrium configuration with fixed central density $\rho_c= 1.5 \times 10^{15}\, \rm g/cm^3$, $\alpha= 0.9$, inner energy density $\rho_{\rm dis}^+= 0.8 \times 10^{15}\, \rm g/cm^3$, and three values of the free parameter $A$. Here and as in the rest of the figures, we have used $\eta= 1.0$ and $\kappa= 100\, \rm km^2$. The solid and dot-dashed curves describe the solutions in the inner core and outer layer of the compact star, respectively. The radius of the phase-splitting surface is represented by the gray vertical lines, which increases as $A$ increases. As expected, it is observed that both the energy density and the pressure decrease with increasing radial coordinate, while the mass is an increasing function with $r$. Note also that both the mass of the dark-energy core as well as the total mass of the star suffer a substantial increase due to the increase in the parameter $A$. We have verified that these configurations satisfy the causality condition throughout the hybrid star.

\begin{figure*}
 \includegraphics[width=8.82cm]{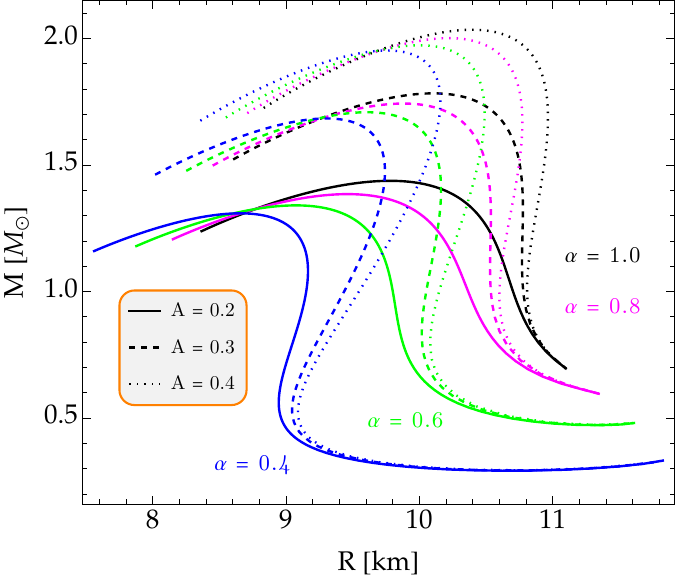}
 \includegraphics[width=8.7cm]{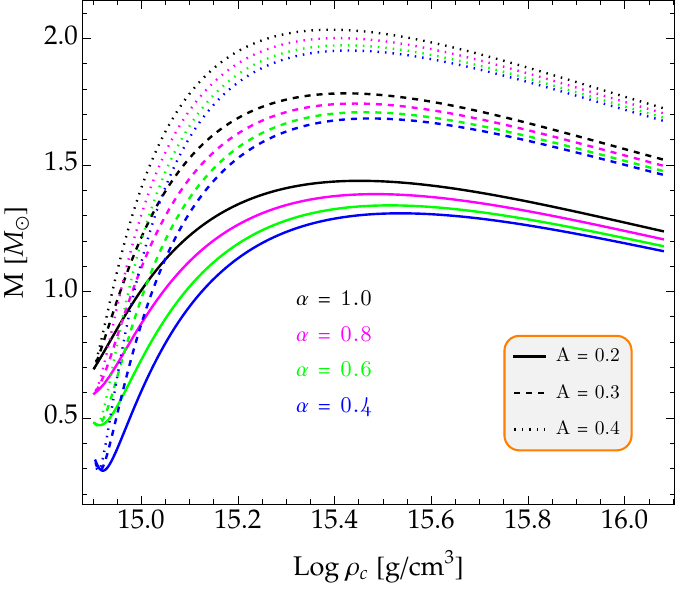}
 \caption{\label{FigMRCdRelations} Mass-radius diagram (left panel) and mass-central density relation (right panel) for hybrid stellar models with EoS (\ref{EoSEq}) for three values of the free parameter $A$: $0.2$ (solid curves), $0.3$ (dashed curves) and $0.4$ (dotted curves). Moreover, the numbers associated with each color indicate the different values of $\alpha= \rho_{\rm dis}^-/\rho_{\rm dis}^+$, where we have considered $\rho_{\rm dis}^+ = 0.8 \times 10^{15}\, \rm g/cm^3$. One can observe that, given a fixed value of $A$ and a given central density, the mass decreases as $\alpha$ gets smaller. On the other hand, if we keep $\alpha$ fixed and vary $A$, we see that the mass undergoes a substantial increase as $A$ increases, for a given central density. In fact, it is possible to go beyond $2M_{\odot}$ when $A= 0.4$ and $\alpha \gtrsim 0.8$. Also note that, according to the right plot for $\alpha= 0.4$ and $\alpha= 0.6$, there would be unstable stellar configurations in the low-central-density region since $dM/d\rho_c <0$. }  
\end{figure*}

\begin{figure}
 \includegraphics[width=8.5cm]{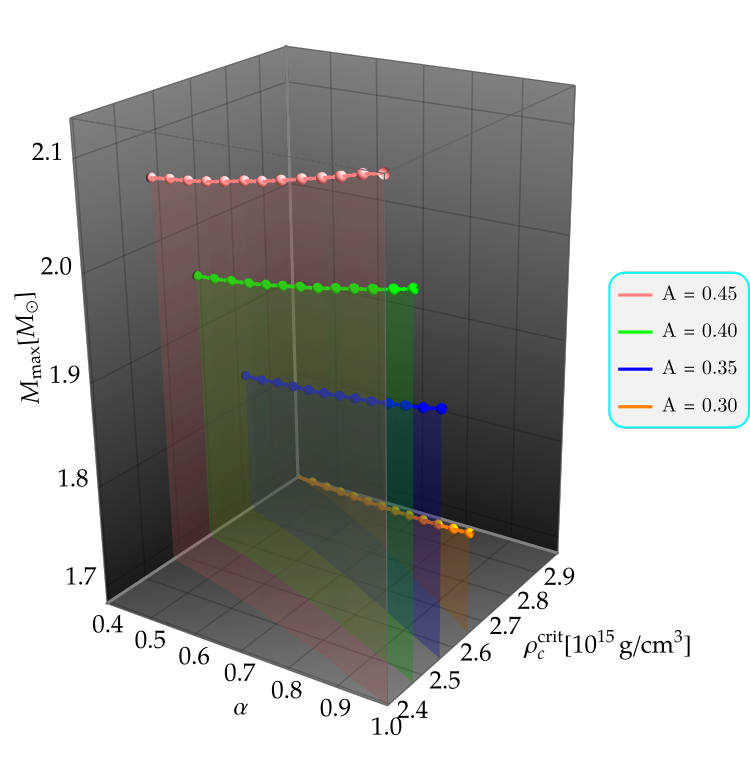}
 \caption{\label{MaxMasses} Maximum mass as a function of the ratio of densities at the interface $\alpha$ and of the critical-central density $\rho_c^{\rm crit}$ for four values of $A$. An increase in $\alpha$ leads to an increasing value of $M_{\rm max}$ but a decreasing value of $\rho_c^{\rm crit}$. }  
\end{figure}

By varying the central density $\rho_c$ it is possible to obtain a family of stellar configurations represented in the well-known mass-radius diagram. This mass versus radius relation is illustrated in the left panel of Fig.~\ref{FigMRCdRelations} for four different values of $\alpha$ and three values of $A$. For a fixed value of $A$, we can see that the maximum-mass values decrease as $\alpha$ gets smaller. Likewise, keeping $\alpha$ fixed, the maximum mass in each curve increases with the increase of $A$. Furthermore, the mass-central density relation is shown in the right plot of the same figure. The mass exhibits a peculiar behavior when $\alpha= 0.4$ (blue lines) and $0.6$ (green curves), namely, the mass first decreases and after reaching a minimum it begins to grow with increasing $\rho_c$. This would indicate that there are two regions of instability according to the necessary but not sufficient condition for stellar stability. In other words, the stable stars would be found only in the region where $dM/d\rho_c> 0$. Nevertheless, a more rigorous analysis of radial stability involves determining the frequencies of the vibration modes when a compact star is radially perturbed. We will return to discuss this in greater detail later.

The central density corresponding to the maximum-mass value is often called the critical-central density $\rho_c^{\rm crit}$, see the maxima on each curve in the right plot of Fig.~\ref{FigMRCdRelations}. Given the range of values for $\alpha \in [0.4, 1.0]$, from Fig.~\ref{MaxMasses} we can easily identify that higher values of $M_{\rm max}$ are obtained for $ \alpha$ larger. However, the critical-central density is less and less as we increase $\alpha$. Remarkably, this qualitative behavior is repeated for any value of $A$. Therefore, according to the standard stability criterion $dM/d\rho_c> 0$, hybrid stars with a dark-energy core stop being stable at a smaller and smaller central-density value with increasing $\alpha$.

\begin{figure}
 \includegraphics[width=8.55cm]{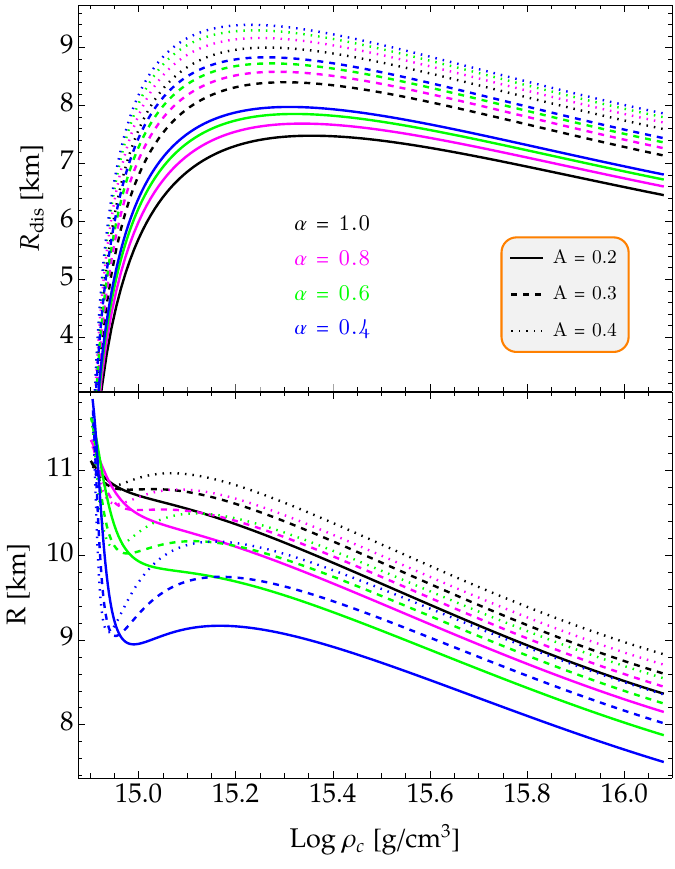}
 \caption{\label{FigRadii} Radius of the discontinuous surface (top plot) and radius of the star surface (bottom plot) as functions of the central energy density for the equilibrium configurations shown in Fig.~\ref{FigMRCdRelations}. A decrease in the values of $\alpha$ results in an increase in $R_{\rm dis}$, for a given $A$. Furthermore, for some combinations of $\alpha$ and $A$, the radius of the star $R$ decreases to a minimum in the low-central-density region, then increases, and after reaching a maximum begins to decrease with increasing central density. }  
\end{figure}

In Fig.~\ref{FigRadii}, we also show the discontinuity radius (upper plot) and the star radius (lower plot) as functions of the central density. For a given $\rho_c$, we observe that the main effect of the parameter $\alpha$ (as it decreases) is a significant increase in the radius of the dark-energy core $R_{\rm dis}$. In the meanwhile, the behavior of the radius of the star $R$ is less trivial with the variation of $\alpha$. For instance, for $\alpha= 1.0$ and $A= 0.2$ (see black solid line), the radius always decreases with increasing $\rho_c$. Nonetheless, when $\alpha= 0.4$ (blue curves), $R$ decreases to a minimum in the low-central-density branch, then increases, and after reaching a maximum starts to decrease again, regardless of the value of $A$.

\begin{figure}
 \includegraphics[width=8.55cm]{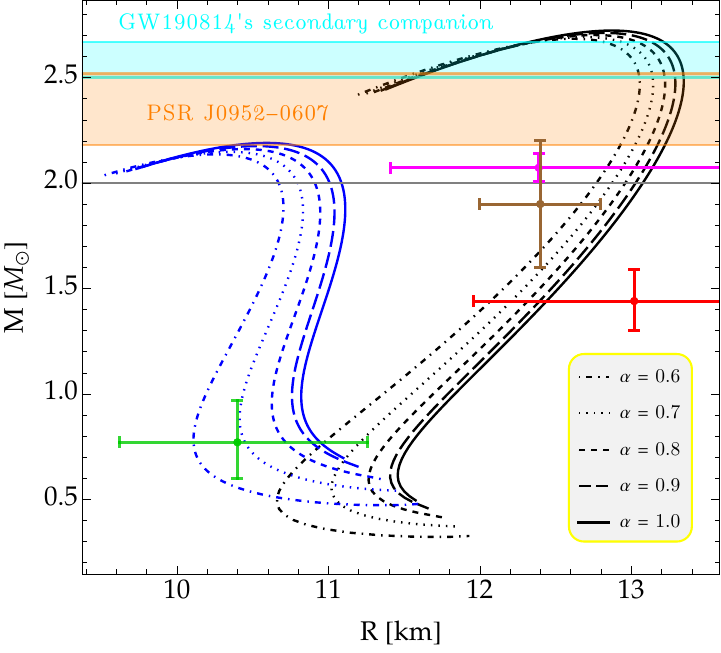}
 \caption{\label{FigMRwithConstraints} Mass-Radius relations of neutron stars with a dark-energy core for $A= 0.48$ and several values of $\alpha$. In addition, we have considered $\rho_{\rm dis}^+= 0.5 \times 10^{15}\, \rm g/cm^3$ (black lines) and $\rho_{\rm dis}^+= 0.8 \times 10^{15}\, \rm g/cm^3$ (blue lines). The gray horizontal streak at $2M_\odot$ represents the two massive NS pulsars J1614-2230 \cite{Demorest2010} and J0348+0432 \cite{Antoniadis2013}. The filled orange and cyan bands stand for the masses of the fastest known spinning NS in the disk of the Milky Way (namely, the pulsar PSR J0952-0607 \cite{Romani2022}) and of the secondary companion detected by the gravitational-wave signal GW190814 \cite{Abbott2020}, respectively. The magenta, brown, red and green dots with their respective error bars represent the masses of the millisecond pulsars PSR J0740+6620 \cite{Riley2021}, 4U 1702-429 \cite{Nattila2017}, PSR J0030+0451 \cite{Miller2019} and supernova remnant HESS J1731-347 \cite{Doroshenko2022}, respectively. The frequencies for the fundamental oscillation mode of these compact stars under rapid phase transitions are shown in Fig.~\ref{FigFrequenConstraints}. }  
\end{figure}

In addition, it is to be expected that our theoretical calculations will be able to describe some compact stars observed in the Universe, such as millisecond pulsars \cite{Demorest2010, Antoniadis2013, Riley2021, Nattila2017, Miller2019, Romani2022} and other compact objects that are still of unknown nature \cite{Abbott2020, Doroshenko2022}. In that regard, Fig.~\ref{FigMRwithConstraints} displays the mass-radius relations compatible with observational measurements for $A= 0.48$, two values of $\rho_{\rm dis}^+$ and several values of $\alpha$. When $\rho_{\rm dis}^+= 0.5 \times 10^{15}\, \rm g/cm^3$ (see black lines), our theoretical predictions consistently describe the pulsar PSR J0952-0607, the fastest known spinning neutron star (NS) in the disk of the Milky Way \cite{Romani2022}. Note also that it is possible to obtain masses above $2.5\, M_\odot$ that are compatible with the secondary component in the GW190814 event, the signal of a compact binary coalescence with the most unequal mass ratio yet measured with gravitational waves \cite{Abbott2020}. Moreover, it can be observed that the Bayesian estimations for the massive pulsar PSR J0740+6620 \cite{Riley2021} and the NS in 4U 1702-429 \cite{Nattila2017} are in agreement with our numerical results.

Since unusually heavy or light neutron stars are the subject of scientific intrigue, Doroshenko and collaborators have analyzed the central compact object within the supernova remnant HESS J1731-347 \cite{Doroshenko2022}. Based on modelling of the X-ray spectrum and a robust distance estimate from Gaia observations, they estimated the mass and radius of such an object to be $M= 0.77_{-0.17}^{+0.20}\, M_\odot$ and $R= 10.4_{-0.78}^{+0.86}\, \rm km$, respectively. This estimate has been represented by the green bars in Fig.~\ref{FigMRwithConstraints}. Furthermore, the authors conjectured that this object is either the lightest NS known, or a ``strange star'' with a more exotic EoS. Nonetheless, it has been argued that the minimum possible mass of a remnant NS is $1.17\, M_\odot$ \cite{Suwa2018}. In that regard, we consider the possibility of describing the central object in HESS J1731-347 as a NS with a dark-energy core when $\rho_{\rm dis}^+= 0.8 \times 10^{15}\, \rm g/cm^3$, see blue curves for the different values of $\alpha$.

\subsection{Radial pulsations}

To investigate the radial stability of the hybrid configurations shown in Figs.~\ref{FigMRCdRelations} and \ref{FigMRwithConstraints}, it is necessary to determine the frequencies of the radial vibration modes. With this in mind, we seek to answer the question: Is the classical stability criterion $dM/d\rho_c >0$ still compatible with radial pulsation analysis for neutron stars with a dark-energy core? As we will see later, this will depend on the type of phase transition.

\begin{figure*}
 \includegraphics[width=8.8cm]{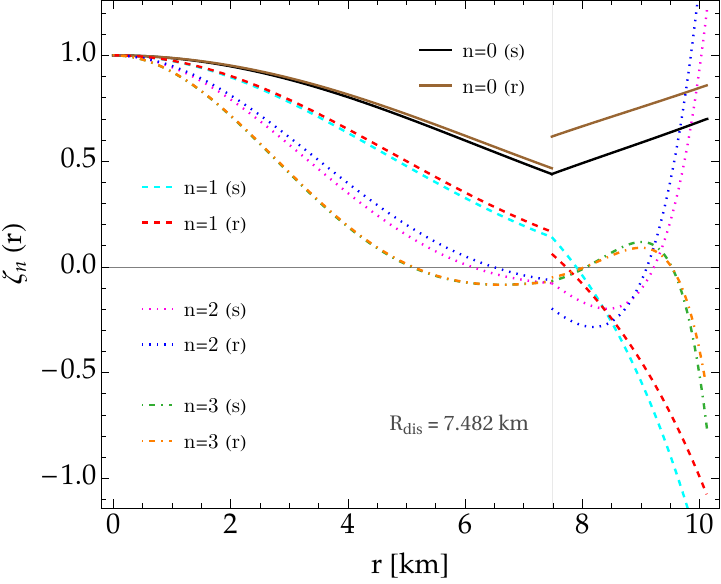}
 \includegraphics[width=8.71cm]{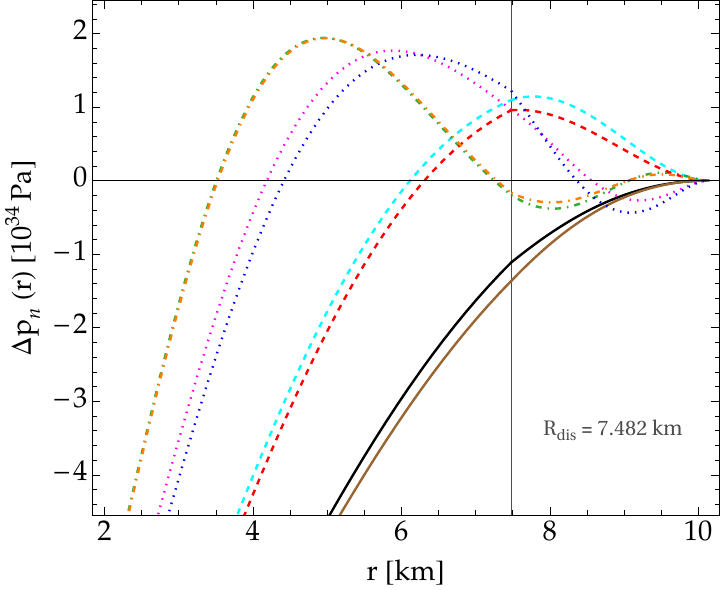}
 \caption{\label{FigEigenmodes} First four vibration eigenmodes $\zeta_n$ (left panel) and $\Delta p_n$ (right panel) within a oscillating hybrid configuration with $\rho_c= 1.5 \times 10^{15}\, \rm g/cm^3$, $\alpha= 0.8$, $\rho_{\rm dis}^+= 0.8 \times 10^{15}\, \rm g/cm^3$ and $A= 0.2$, leading to $R= 10.144\, \rm km$ and $M= 1.235\, M_\odot$. The numerical solution corresponding to the $n$th normal pulsation mode contains $n$ nodes between the center and the surface of the star, and the letters ``s'' and ``r'' in parentheses represent the slow and rapid phase transition, respectively. Moreover, the vertical line indicates the radius of discontinuity, where the eigenfunctions $\zeta_n$ jump due to the boundary condition in the case of rapid phase transitions (\ref{JuncCond2}). }  
\end{figure*}

Given a specific value of central density $\rho_c$, we integrate the radial oscillation equations (\ref{ROEq1}) and (\ref{ROEq2}) with the corresponding boundary conditions (\ref{BCforRO1}) and (\ref{BCforRO2}) from the center up to the surface of the star. Nonetheless, note that at the interface we must use the junction conditions (\ref{JuncCond1}) and (\ref{JuncCond2}) for the slow and rapid phase transition, respectively. Similar to the case of single-phase compact stars, the integration is carried out for a set of trial values $\omega^2$ and the appropriate eigenfrequencies are those that fully satisfy the boundary conditions. For example, for a hybrid configuration with $\rho_c= 1.5 \times 10^{15}\, \rm g/cm^3$, $\alpha= 0.8$, $\rho_{\rm dis}^+= 0.8 \times 10^{15}\, \rm g/cm^3$ and $A= 0.2$, Fig.~\ref{FigEigenmodes} shows the first four vibration eigenmodes $\zeta_n$ and $\Delta p_n$, where the $n$th normal mode contains $n$ nodes between the origin and the surface. Nodes are places inside the star where the displacement is always zero, so that the pulsation mode corresponding to $n=0$ (known as the fundamental mode) has no nodes, the first overtone ($n=1$) has one node, and so on. Moreover, we have considered normalized eigenfunctions $\zeta(0)= 1$ at the center, and the letters ``s'' and ``r'' in parentheses stand for the slow and rapid phase transition, respectively. As expected, the perturbations $\zeta_n$ undergo jumps at the discontinuity radius due to the junction condition in the case of rapid phase transitions (\ref{JuncCond2}). The smaller the mode, the greater the jump. Meanwhile, the perturbations $\Delta p_n$ are continuous for both phase transitions.

\begin{figure*}
 \includegraphics[width=8.7cm]{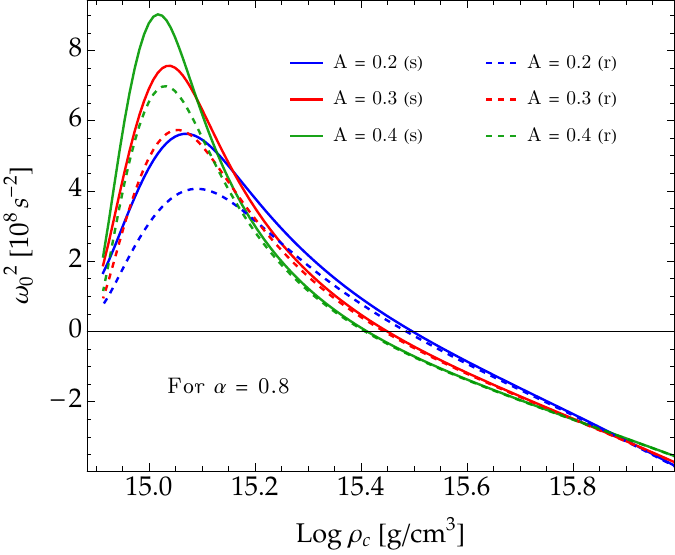}
 \includegraphics[width=8.806cm]{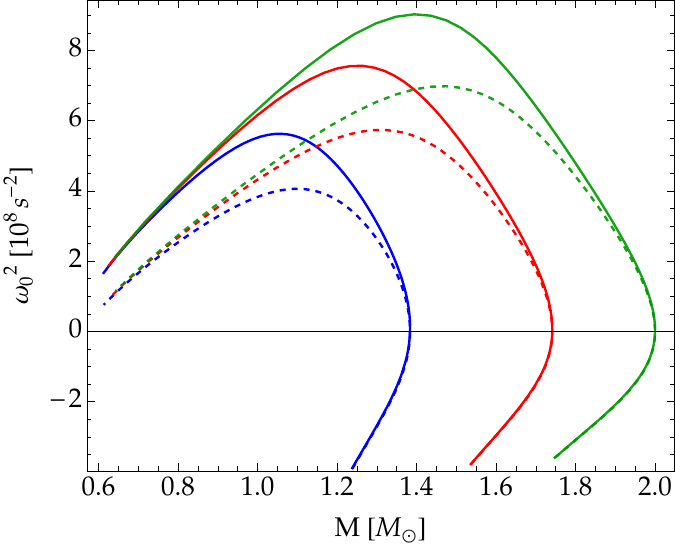}
 \caption{\label{FigFrequen08} Left panel: Squared frequency of the fundamental oscillation mode versus central density for hybrid stellar models with EoS (\ref{EoSEq}) by using $\rho_{\rm dis}^+= 0.8 \times 10^{15}\, \rm g/cm^3$, three values of $A$ and $\alpha= 0.8$ for both slow (solid lines) and rapid (dashed lines) phase transitions. In the low-central-density branch, the fundamental mode frequencies for slow phase transitions are higher than those for rapid transitions. Right panel: Squared frequency of the fundamental mode as a function of the total gravitational mass, where it can be observed that the analyzed configurations stop being stable at the maximum-mass point. }  
\end{figure*}

\begin{figure*}
 \includegraphics[width=8.6cm]{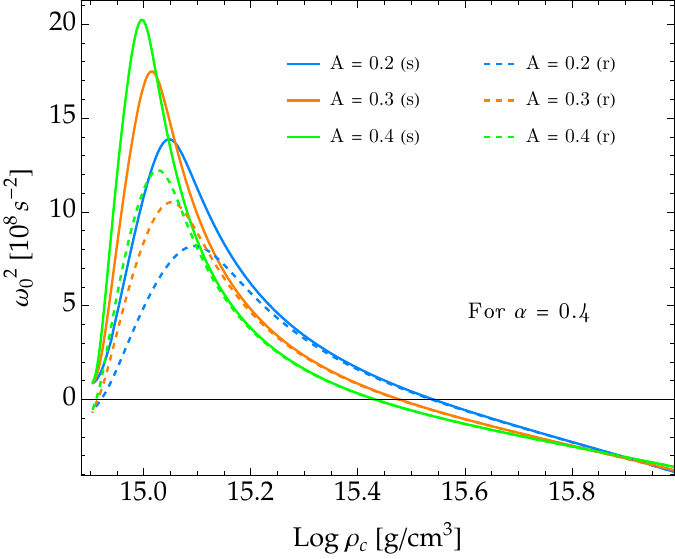}
 \includegraphics[width=8.865cm]{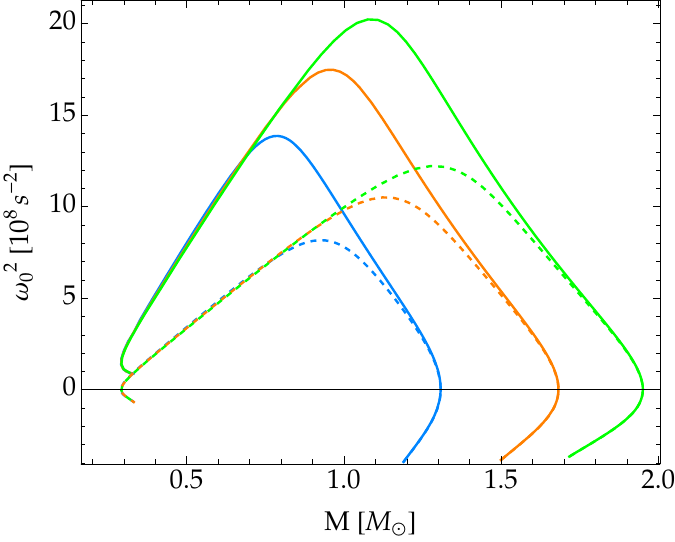}
 \caption{\label{FigFrequen04} Squared frequency of the fundamental vibration mode as a function of the central density (left panel) and of the gravitational mass (right plot) as in Fig.~\ref{FigFrequen08}, with the only difference that we have now used $\alpha= 0.4$ in our numerical calculations. Notably, only rapid phase transitions are capable of predicting the unstable branch at low central densities. See also the right plot of Fig.~\ref{FigMRCdRelations} to identify the configurations that lie in the first region where $dM/d\rho_c <0$ for $\alpha= 0.4$. }  
\end{figure*}

Notice that, the numerical solution of the radial pulsation equations gives an infinite discrete set of eigenvalues $\omega_0^2 < \omega_1^2 < \cdots $, where the stable configurations correspond to $\omega_n^2>0$. Therefore, it is enough to determine $\omega_0^2$ to know if a hybrid star is stable or not with respect to radial perturbations. Figures \ref{FigFrequen08} and \ref{FigFrequen04} illustrate the behavior of the squared frequency of the fundamental vibration mode as a function of central density and total gravitational mass for $\alpha= 0.8$ and $\alpha= 0.4$, respectively. Given a specific value of $\alpha$, all values of $A$ yield similar behaviors, that is, the squared frequency of the fundamental mode increases until reaching a maximum value and then decreases with increasing $\rho_c$ for both slow and rapid phase transitions. For a fixed value of $A$, the fundamental mode frequencies for slow phase transitions are higher than those for rapid transitions. This behavior is similar to the scenario of neutron stars composed by a quark-matter core and hadronic-matter crust \citep{Pereira2018}, including the case of inverted hybrid stars \citep{Zhang2023}. One can observe that for a larger parameter $A$ the hybrid star has a smaller central density by which its squared eigenfrequency $\omega_0^2$ reaches a maximum value. Besides, remark that the radial stability ceases (when $\omega_0^2=0$) at a smaller and smaller value of $\rho_c$ as $A$ increases. In other words, an increase in $A$ gives rise to a reduction in the stability of neutron stars with a dark-energy core.

Maximum-mass values coincide with $\omega_0^2=0$ regardless of the value of $A$. This means that after the maximum-mass configuration we have $\omega_0^2< 0$ (i.e., the frequency is purely imaginary) and hence the stars will undergo gravitational collapse when $\rho_c > \rho_c^{\rm crit}$. Notwithstanding, we must point out that, mainly in the low-central-density region, the fundamental mode frequencies obtained by rapid phase transitions differ significantly from slow transitions. Even more interesting is the case when $\alpha= 0.4$, where unstable stars should exist in the low-mass region according to the blue curves in Fig.~\ref{FigMRCdRelations}. Slow phase transitions (solid curves in Fig.~\ref{FigFrequen04}) are unable to predict such unstable configurations, while the rapid phase transitions (represented by dashed lines) indicate a small unstable branch for low mass stars. As a consequence, we can conclude that only rapid phase transitions are compatible with the standard stability criterion $dM/d\rho_c >0$ in the sense that $\omega_0^2$ is exactly zero at a central-density value corresponding to $dM/d\rho_c =0$ and, therefore, there are two regions of instability for sufficiently small $\alpha$.

\begin{figure*}
 \includegraphics[width=8.695cm]{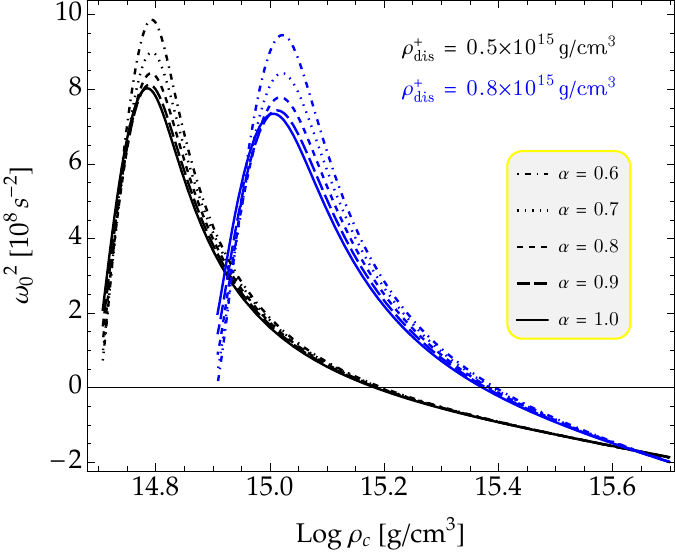}
 \includegraphics[width=8.80cm]{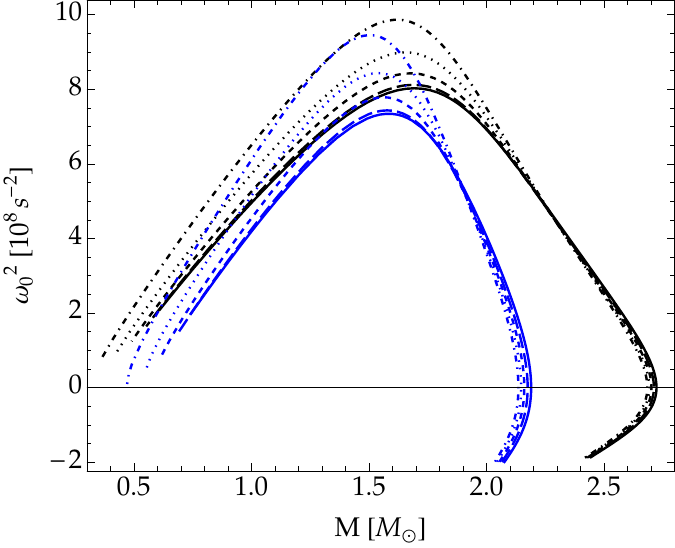}
 \caption{\label{FigFrequenConstraints} Oscillation spectrum under the effect of rapid phase transition for the stellar configurations presented in Fig.~\ref{FigMRwithConstraints}, where we have used a fixed value of $A= 0.48$ and five values of $\alpha$. Note also that the black lines correspond to the inner energy density at the interface $\rho_{\rm dis}^+= 0.5 \times 10^{15}\, \rm g/cm^3$, while the blue curves represent the case where $\rho_{\rm dis}^+= 0.8 \times 10^{15}\, \rm g/cm^3$. Regardless of the value of $\alpha$ and $\rho_{\rm dis}^+$, it can be observed that the maximum-mass point can be used as a turning point (where the stars stop being stable) since from there we obtain $\omega_0^2< 0$. }  
\end{figure*}

Under the effect of rapid phase transitions, in Fig.~\ref{FigFrequenConstraints} we have calculated the squared frequency of the fundamental pulsation mode for the equilibrium configurations shown in Fig.~\ref{FigMRwithConstraints}. The comparison between the black and blue curves indicates that an increase in $\rho_{\rm dis}^+$ leads to an increasing critical-central density where the stars are no longer stable. This means that a large $\rho_{\rm dis}^+$ favors the stability of neutron stars with a dark-energy core. Besides, regardless of the value of $\alpha$ and $\rho_{\rm dis}^+$ in the right plot of Fig.~\ref{FigFrequenConstraints}, one sees that the maximum mass can be used as a turning point (where the stars stop being stable) since from there we obtain $\omega_0^2< 0$. Therefore, our results show that the existence of neutron stars with a dark-energy core is possible in the Universe for two simple reasons: They are dynamically stable (at least until before reaching the maximum-mass point) and are consistently compatible with the observational measurements.

\section{Concluding remarks}\label{conclusion}

Within the context of Einstein gravity, we have investigated the equilibrium structure of hybrid stars, where the internal core contains a dark-energy fluid while the external layer is ordinary matter. In other words, we extended the investigation of compact stars by confining dark energy in the star's core and by adding a common-matter crust on the core. The dark-energy core was considered to be made of a negative pressure fluid ``$-B/\rho$'' plus a barotropic component ``$A\rho$'', namely the Chaplygin-like EoS, which drives the accelerated expansion of the Universe. In the meanwhile, the outer layer of the compact star was described by a well-known polytropic EoS. The effect of the rate of energy densities at the phase-splitting surface $\alpha= \rho_{\rm dis}^-/\rho_{\rm dis}^+$ on the macroscopic properties has been investigated in detail.

We found that the resulting mass-radius diagrams are substantially different from dark energy stars without an ordinary-matter crust. In particular, previous studies \cite{Panotopoulos2020EPJP, Pretel2023EPJC} showed that the radius of dark energy stars with a single phase decreases as the mass becomes smaller in the low-central-density region. Nonetheless, our results revealed that, for small values of $\alpha$ (such as $\alpha= 0.4$), it is possible to obtain large radii, similar to the scenario of compact stars made of pure hadronic matter. Remarkably, some of these high-radius and low-mass stellar configurations correspond to a branch where $dM/d\rho_c <0$. We have also noticed that both the mass of the dark-energy core as well as the total mass of the neutron star suffer a substantial increase due to the increase in the parameter $A$. On the other hand, for a fixed value of $A$, the maximum-mass values on the $M(\rho_c)$ curve decrease as $\alpha$ gets smaller. Additionally, for a given $\rho_c$, we observed that the main consequence of the parameter $\alpha$ (as it decreases) is a significant increase in the radius of the dark-energy core $R_{\rm dis}$.

We examined in detail the effect of dark energy on the normal vibration modes of a hybrid star. To that end, we have adopted two different approaches for the junction conditions due to phase transitions at the discontinuity radius. We have found out that, for a larger parameter $A$, the hybrid star has a smaller value of central density by which its squared eigenfrequency $\omega_0^2$ reaches a maximum value. Indeed, an increase in $A$ gives rise to a reduction in the stability of neutron stars with a dark-energy core. Note that the value of $A$ has an important consequence on the EoS parameter $B$ through Eq.~(\ref{EqB}), so that the term responsible for the accelerated expansion of the Universe (that is, ``$-B/\rho$'') strongly depends on the choice of $A$. Furthermore, the fundamental mode frequencies obtained by rapid phase transitions differ significantly from slow transitions mainly in the low-central-density region. Our results revealed that only rapid phase transitions are compatible with the standard stability criterion $dM/d\rho_c >0$ in the sense that $\omega_0^2$ is exactly zero at a central-density value corresponding to $dM/d\rho_c =0$ and, therefore, there are two regions of instability for sufficiently small $\alpha$. However, this statement no longer holds for slow phase transitions.

The most interesting phenomenological cases, compatible with the observational mass-radius constraints and respecting the causality condition, have been obtained for $\rho_{\rm dis}^+= 0.5 \times 10^{15}\, \rm g/cm^3$ and $\rho_{\rm dis}^+= 0.8 \times 10^{15}\, \rm g/cm^3$ with $A= 0.48$ and a wide range of values for $\alpha$. Consequently, our work has shown that the existence of neutron stars with a dark-energy core is possible in the sense that they are dynamically stable under small radial perturbations and are compatible with the recent astronomical measurements.

\begin{acknowledgments}
JMZP acknowledges financial support from the PCI program of the Brazilian agency ``Conselho Nacional de Desenvolvimento Cient{\'i}fico e Tecnol{\'o}gico''--CNPq. MD thanks Coordenação de Aperfeiçoamento de Pessoal de Nível Superior (CAPES) -- Finance code 001, CNPq under Grant 308528/2021-2, and Fundação de Amparo à Pesquisa do Estado de São Paulo (FAPESP) under Thematic Project No.~2017/05660-0 and Grant No.~2020/05238-9. SBD thanks CNPq for partial financial support. This work has been done as a part of the Project INCT-Física Nuclear e Aplicações, Project number 464898/2014-5.
\end{acknowledgments}\


%

\end{document}